# Impact of Autonomous Vehicle Technology on Long Distance Travel Behavior

**Abstract ID: 782900**

**Maryam Maleki, Yupo Chan, and Mohammad Arani**
**University of Arkansas at Little Rock**
**2801 S University Ave, Little Rock, AR 72204, USA**

## Abstract

Although rapid progress in vehicle automated technology has sped up the possibility of using fully automated technology for public use, little research has been done on the possible influences of autonomous vehicles (AVs) technology on long distance travels. This technology has the potential to have a significant effect on intercity trips. This study analyzed a travel survey to anticipate the impact of this technology on long distance trips. We have divided trips into two different categories including trips for pleasure and trips for business. Different hypotheses based on the authors knowledge and assisted by existing literature have been defined for each type of trips. By using the Pearson method these hypotheses have been tested and the positive or negative responses from respondents have been evaluated. The findings show that using AVs for pleasure trips can increase the number of travelers and stimulate people to choose longer distances for their trips. In addition, people enjoy more and will be interested to travel more frequently. For business trips AV technology can reduce the travel costs, and job-related stress. Unlike pleasure trips for which people are not interested in traveling at night, business travelers prefer to travel at night.

## 1. Introduction

In recent years, different aspects of human life such as the transportation system have been revolutionized by various advanced technologies. The transportation system as one of the important sectors in today's life has been meaningfully influenced by Information Communication Technology (ICT) development. One of the technological advancements in automation that is changing very rapidly is autonomous vehicle technology. Numerous companies are involved in conducting different experiments in Level 4 of autonomous vehicles as defined by the Society of Automotive Engineers (SAE), which includes vehicles that are autonomous under specific conditions. These efforts can lead to the growing expectations from autonomous vehicles in the world [1]. One of the most important goals of autonomous vehicles is to improve safety. Beside safety some other aspects such as improving traffic flow efficiency, convenience, and capacity as well as reducing congestion are expected to be fulfilled by automation. By means of different technologies such as cameras, GPS, sensors, Adaptive Cruise Controls (ACC), advanced system of assisting drivers, AV can drive itself without the participation of the any driver and in this situation human drivers will be set free [2]. Generally, using AV all users can be freed as passengers inside the car. Consequently, passengers can allocate their travel time to different actions such as reading, eating, watching movies, playing games, sleeping, working, enjoying looking at the environment and so forth. Therefore, the mentioned abilities of AV technology can be a proper motivation for people to use AVs instead of conventional ones [3][4]. Evaluating the potential influences of AVs on long distance trips is one important area in which insignificant research has been done to date. This kind of trip is one of the important areas of travel growth and are significantly more affected by travel time and cost than short distance travels [5]. Based on the points mentioned above, AVs have the potential to significantly alter the way long distance travelers react to this advanced technology. To be more precise, travelers who are on business trips have the same opportunity to work during their route as they would on a plane, but they can save more time because of omitting airport wait times and mode change inconveniences to reach their final destination. In addition, people who travel for pleasure are able to enjoy more quality time and personal time. Also, while traveling they can stay in contact with advanced communication devices with less cost than airlines and high-speed trains [6].

## 2. Background

### 2.1. Emergence of autonomous vehicles technology

In recent years, AV technology has advanced rapidly. AVs, also known as self-driving cars, can be an effective technology [7] [8] which may have different benefits such as reducing traffic congestion, increasing the productivity of using travel time, diminishing crashes, enhancing fuel efficiency, reducing emissions, and parking advantages, and freedom for drivers [4] [9] [10] [11]. Different kinds of automation facilities including self-parking and automatic braking system have already been used in some vehicles. Google and various auto manufacturers have set goals to commercialize AVs by the end of this decade [12]. Aside from SAE, five levels (Level 0 to Level 4) of automation has been identified by NHTSA as well. Level 0 belongs to the vehicles that all functions of the car is controlled by human driver while Level 4 depends on complete vehicle control when the car can drive automatically without the interference of human driver. In level 4, there is a great potential to influence long distance travel behavior [6]. In four states in the US as well as Washington DC, the process of testing the AV on highways was legislated in 2014 [13]. Due to swift gains in the AV technology in recent years, there is a great possibility of sales of vehicles which are partially or completely automated to the public. It has been predicted that AVs can form between 2% to 5% of all the new cars sold in the 2020s, even though the first cars will be available just to those who can afford the cost. Moreover, there is an uncertainty about the future of market penetration of AVs. While some people like Litman (2017) [14] anticipate that the market penetration in the U.S. will be more than 50 % in the 2025s others like Bierstedt (2014) [12] predict a later adaption. They estimate the 50% market penetration between 2035 and 2050. Although there is uncertainty about the pace of the transition, there is a general consensus that AV technology will be available in the future and it will influence people's travelling behavior as well as transportation system.

### 2.2. Long-distance travel

Extensive population growth as well as broad economic development have caused continuous increase in travel demand, average travel distance, and consequently resulting in more traffic delays and congestion. Average daily person miles traveled (PMT) in the US, for instance, increased from 83.1 miles to 90.4 miles between 2001 and 2009. Meantime, average vehicle miles traveled (VMT) increased from 49.8 miles to 54.4 miles between 1990 and 2009. These concerns have aroused researchers to acknowledge the needs of a new approach that can address current congestion problems in addition to growing travel demand in the future [15]. In the United States, trips that are longer than 50 mi (80 km) account for 2% of the trips but they are about 23% of vehicle miles traveled [16]. The amounts are equivalent in the United Kingdom and trips longer than 50 mi (80 km) accounts for 2.3% of trips but 30.2% of the whole distance traveled. Hence, long distance trips are clearly a significant travel market for intercity highway projects, air travel, or high-speed rail demand [17]. Reliable foresight of intercity travel can be a momentous tool for planning long-distance traveling transportation improvements, such as intercity highway and various transit projects.

### 2.3. Convergence

There is a growing body of literature that evaluate the influence of emerging AVs on travel behavior to understand the travelers' reaction to this advanced and novel mode. Studies showed that this new technology has the potential ability to make changes in the market structure and travelers' decision on choosing their mode of travel [3] [18]. Previous studies mostly focused on the classic features like travel time and travel costs to evaluate the travelers' preference between various AV adaptation forms including "private automated vehicles (PAVs) and shared automated vehicles (SAVs)" [19] [20]. Other characteristics in addition to travel costs and travel time, such as travel distance, in the travelers" mode choice have been evaluated in the study by Arentze et al. [21]. In some studies just short distance travels have been evaluated [18] [19]. In some other studies authors have shown that other features such as psychological characteristics and attitude have the potential to influence people's inclination to use AVs [18]. [20] [22]. The purpose of travel can be another important factor that can influence travelers' decision in travel behavior [23] [24]. In some research it is stated that there is an uncertainty about AV acceptance and the effect of this technology on travel behavior [3] [22]. LaMondia et al. [6] have developed a survey and after that a simulation to predict the influence of AVs on long-distance travel mode choices for various trip distance and purposes. They demonstrated that

*Maleki, Chan, and Arani*travel distance plays an important role on travelers' preference for selecting between AVs and airplane. In addition, the findings of their research showed that compared with travel time, monetary costs are less important. Ashkrof et al. [25]employed a survey in the Netherlands to evaluate the influence of AV on travel mode choice. In their research respondents had to choose among public transportation, ordinary cars, and autonomous cars for various distances and travel purposes. Their study showed that travelers have different inclinations for various travel distance and purposes. In another research, authors showed that AVs can cause a change of destination choice. Also by using AVs people are more interested in longer trips for personal vehicles [26]

## 3. Method

It is clear that there are many different findings on the acceptance of AVs on long distance travel, and there is a need to sort out these differences. Based on the above literatures and their personal experiences different hypotheses have been postulated by the authors. As AV is a technology of the future, solid data is not available to delineate the differences in previous findings. In this research, we resort to a survey. In short, information is gathered through such a questionnaire by way of an online survey that was distributed among experts in the transportation field in the U.S. The respondents of the questionnaires are experts in the transportation field and they are completely familiar with AV technology. We tried to explain about this technology to respondents and make them familiar with the subject before distributing the questionnaires. The data of 66 respondents were received in this study. We have divided trips based on their purpose into two categories, namely, Business trips and Pleasure trips. We have used a five-choice Likert spectrum which is one of the most common measurement scales in a questionnaire. The socioeconomic characteristics of the respondents are shown in Table 1.

**Table 1:** Characteristics of the respondents

| Characteristic | | Numbers | Characteristic | | Numbers |
|---|---|---|---|---|---|
| Education | Bachelor | 15 | Sex | Man | 45 |
| | Master | 24 | | Woman | 21 |
| | PhD | 27 | | Total | 66 |
| | Total | 66 | Income $ (per year) | Less than 30000 | 4 |
| Age | 20-30 | 3 | | 30,000 - 60,000 | 12 |
| | 30-40 | 25 | | 60,000 – 100,000 | 16 |
| | 40-50 | 23 | | 100,000 – 150,000 | 20 |
| | Above 50 | 15 | | More than 150,000 | 14 |
| | Total | 66 | | Total | 66 |

The following hypotheses have been defined for two different kinds of trips and were tested using the Pearson method.

### 3.1. Hypotheses for pleasure trips

Hypothesis 1 (H0P1): Using AVs will increase the number of "mobility-challenged" people. By "mobility challenged," we are referring to those who are not able to travel, or it is difficult for them to travel, such as the disabled, elders, teenagers who don't have a driver's licenses, and families with children.
Hypothesis 2 (H0P2): AVs will cause travelers to choose longer distances for traveling.
Hypothesis 3 (H0P3): As people using AV technology have this opportunity to rest in the car, they will prefer to travel at night.
Hypothesis 4 (H0P4): AV technology will cause people to enjoy more of their trips.
Hypothesis 5 (H0P5): AV will increase the frequency of peoples' trips.

**Table 2**: Pearson Correlation Coefficient for Pleasure Trips Hypotheses

|    |                     | Number of people | Longer distance | Night travel | Enjoyment | Frequency of trips |
|----|---------------------|------------------|-----------------|--------------|-----------|--------------------|
| AV | Pearson Correlation | 0.732            | 0.683           | -0.336       | 0.812     | 0.572              |
|    | Sig. (2-tailed)     | <0.001           | <0.001          | <0.001       | <0.001    | <0.001             |
|    | N                   | 66               | 66              | 66           | 66        | 66                 |

The results of Table 2 show that the correlation coefficient of number of people and using AV technology is r = 0.732, which shows a significant relationship with a 99% confidence level (P<0.01). Therefore, it can be statistically deduced that there is a considerable positive relationship between using AVs and the number of people. It means that by increasing the use of AVs, the number of people who are willing to travel will increase. Accordingly, the first hypothesis (H0P1) can be substantiated. It means using AVs can increase the willingness of travel for many people who are not able to travel, or it is difficult for them. The disabled, elders, families with kids, people who have a temporary mobility problem, people who are stressed, people who don't like driving or scared of driving and so on will choose AVs and consequently increase the number of travelers. Similarly, we can conclude from the table that using AV technology will cause people to select longer distances for their trips and the second hypothesis (H0P2) can be confirmed also. When people are comfortable during a trip, they are more likely to choose longer distances. Our findings have shown that people preferred daylight traveling overnight traveling for pleasure using an AV. It could be they were stressed during night travels and they preferred to use cars that have human drivers over AVs. In this case, the correlation coefficient of traveling at night is r = - 0.336 that shows inverse relation between using AVs and night travel. Hence, the third hypothesis (H0P3) can be rejected. Moreover, a correlation coefficient of r = 0.812 for enjoyment demonstrate a strong positive relationship between using AVs and the corresponding pleasure in the experience. Not surprisingly, people who travel for pleasure will be more delighted to use this technology because they have time to engage in different activities such as reading, eating, watching movies, taking pictures, and playing games. Therefore, the fourth hypothesis (H0P4) is substantiated. Moreover, the results of the table show that there is a significant positive relationship among using AVs and the frequency of trips. It is statistically inferred that when people use AVs, they are more eager to travel, and they decide to travel more frequently. Consequently, the fifth hypothesis (H0P5) can be confirmed.

### 3.2. Hypothesis for business trips

H0B1: Using AVs will decrease the travel cost because people have more flexible time to travel. For example, they can travel at night and rest in the car during the route without needing to go to a hotel.
H0B2: People who travel by AVs can work during the route.
H0B3: People prefer to travel at night and can rest during the route and will be ready for their business meeting or appointment.
H0B4: People who use AVs will have less travel-related stress during their trip.

**Table 3**: Pearson Correlation Coefficient for Business Trips Hypotheses

|    |                     | Decrease cost | Job meeting | Night travel | Reducing job stress in trip |
|----|---------------------|---------------|-------------|--------------|-----------------------------|
| AV | Pearson Correlation | 0.634         | 0.743       | 0.541        | 0.498                       |
|    | Sig. (2-tailed)     | 0.001         | 0.000       | 0.000        | 0.001                       |
|    | N                   | 66            | 66          | 66           | 66                          |

Table 3 shows that the correlation coefficient between decreasing cost of travel and using AVs technology is r = 0.634, which demonstrates a significant relationship with a 99% confidence level (P<0.01). Therefore, it can be statistically deduced that there is a considerable positive relationship between using AVs and decreasing cost. This means that the first hypothesis (H0B1) can be supported. To explain more, when people travel for business, they have the opportunity to rest in the car without a need to book a hotel, and they can go to their meeting or business appointment being refreshed. In addition, for a correlation coefficient of r = 0.743 it means that there is a positive relationship between using AVs and working in a driverless car. This suggests that the second hypothesis (H02B) can be confirmed. It shows that people are willing to do different activities related to their job in the car as there is no need to be involved in driving. This means that in addition to a physical meeting, they can handle their meetings through skype calls or any other kinds of digital meeting methods. Unlike pleasure travel, people who are traveling for business prefer to travel at night. The findings shown in Table 2 (r = 0.541) shows a moderate relationship between using AVs and traveling at night and this substantiated the hypothesis. For example, when a person has an important meeting in the early morning, they can travel at night and rest on the way. The main goal of this type of trip is business so they can perform their job duties with comfort during their route. In addition, the findings shown in table 2 reveal that it can be statistically inferred there is a significant relationship between reducing travel related stress and using AVs. This translates into the fact that H0B4 can be confirmed. When people are freed from driving they are more likely to be ready for their job because their minds are not as likely to be occupied by so many different distractions about driving and they will be relaxed and energetic when they arrive to the destination.

## 4. Conclusion

Nowadays self-driving cars technology has experienced significantly rapid advances and cars that are completely automated will happen in near future. This technology may have significant impact on a traveler's behavior. One of the areas that might experience significant changes is long distance travel-a subject that little study has been done-. Long distance travel is one of the most important travel developments in the United States and can be changed dramatically by AV technology. This study analyzed the impact of this technology on intercity travels by testing different hypotheses. By way of surveying transportation experts these hypotheses have been either substantiated or rejected. Findings of this study demonstrates that using AVs technology may result in different changes in travel behavior. This technology in pleasure trips, will increase the number of travelers. People who were not able to travel or it was so difficult for them to take trip will be motivated to travel more. In addition, AV technology will cause travelers to travel longer distances as they are so comfortable during the trip and can allocate their time to so many activities rather than driving. People traveling on pleasure are not interested in traveling at night because they may feel nervous using AVs at darkness. For business trips people are able to reduce their cost when they do not need to rest in hotel as they can do so in the AV. Moreover, this technology can reduce travel-related stress because travelers do not need to concentrate on driving and can be more comfortable during the trip. Our findings show that when people travel for business, they are more likely to travel at night. People also have the opportunity to work during the route and can have their physical meeting with their colleagues in the AVs as well as meetings using the internet.